\documentclass[twocolumn]{revtex4}
\usepackage{graphicx}
\usepackage{amssymb}
\newcommand{\beq}{\begin{equation}}
\newcommand{\enq}{\end{equation}}
\newcommand{\bea}{\begin{eqnarray}}
\newcommand{\ena}{\end{eqnarray}}
\newcommand{\ad}{a^{\dag}}
\begin{document}

\title{Mott transition in anharmonic confinement}
\author{Emil Lundh}
\affiliation{Department of Physics, Ume{\aa} University, 
SE-90187 Ume{\aa}, Sweden}
\begin{abstract}
Two effects are identified that affect the visibility of the 
Mott transition in an atomic gas in an optical lattice confined 
in a power-law potential. The transition can be made more 
pronounced by increasing the power law, but at the same time, 
experimental uncertainty in the number of particles will 
induce corresponding fluctuations in the measured condensate 
fraction. Calculations in two dimensions 
indicate that a potential slightly more flat-bottomed than a 
quadratic one is to be preferred for a wide range of 
particle number fluctuation size.
\end{abstract}
\maketitle
In a spectacular experiment a few years ago, the transition from a 
superfluid to a Mott insulator was realized in a gas of cold atoms 
in an optical lattice by Greiner {\it et al.} \cite{greiner2002}. 
The experiment marked the birth of a new subfield of physics, 
where theoretical statistical-physics models, until then considered 
to be mostly of academic interest or crude approximations to 
real materials, can now be realized in experiment. With the help 
of atoms in optical lattices, it is hoped that fundamental 
questions regarding phase transitions can be addressed, as well as 
applications to e.~g.\ quantum information.
Nevertheless, a few practical issues remain to be sorted out. One such 
issue concerns the role of the confining potential. 

A gas of spinless bosonic atoms in an optical lattice 
is known to be well described by the Hubbard model \cite{jaksch1998}, 
\beq\label{hamiltonian}
H = \frac12 \sum_{r}\ad_r\ad_r a_r a_r -
\frac{t}{2} \sum_{<rr'>}\ad_r a_{r'}
+\sum_{r}[V(r)-\mu] \ad_r a_r,
\enq
where $t$ is the tunneling matrix element, $\mu$ is the chemical 
potential, $V(r)$ is the spatially dependent trapping 
potential, and $r$ is a dimensionless site index. The operators
$a_r$,$\ad_r$ destroy and create a particle at the site $r$,
respectively, and obey Bose commutation relations. The units are 
here chosen so that the on-site interaction strength, i.~e., the 
prefactor of the first term in (\ref{hamiltonian}), 
is unity. The sum
subscripted $<\!rr'\!>$ runs over all pairs of nearest-neighbor sites.
In the absence of an external potential $V(r)$, this Hamiltonian 
exhibits a quantum phase transition at zero temperature, 
separating two ground states \cite{sachdev1999}: When the tunneling 
matrix element $t$ is strong 
enough, there is phase coherence over the entire sample which 
puts the system in its superfluid state, and there exists an 
accompanying Bose-Einstein condensed fraction of the gas that 
can be measured in time-of-flight experiments \cite{stoferle2004}. 
For weaker tunneling, phase coherence is lost, the number of 
atoms per site is locked to an integer, and number fluctuations 
are suppressed. This is the Mott insulating state. 

In the experiment reported in Ref.\ \cite{greiner2002}, 
just as in nearly all experiments on optical lattices, the atoms 
were contained in a quadratic, i.~e., harmonic-oscillator 
potential, in order for them 
not to escape. As a result, the transition 
between superfluid and Mott insulator is not simultaneous over the 
whole sample, but takes place via an intermediate state where part of 
the atomic cloud is superfluid and part is Mott insulating. Such a 
state is often called a ``Mott plateau'' state because of the 
characteristic density profile, in which the density is fixed to an 
integer in confined regions. This is well known and it is also  
well understood \cite{bergkvist2004,batrouni2002}. In effect, it 
is the local confining potential that 
contributes a (negative) addition to the chemical potential, 
resulting in a spatially dependent critical point for the 
phase transition. 

In order to see a clear Mott transition one would need to get 
rid of the effects of the confining potential. It has therefore been 
proposed that the harmonic-oscillator potential could be replaced by a 
more flat-bottomed variety, e.~g., a quartic or sixth-power potential, 
more similar to a square well \cite{gygi2006}. In such a potential, 
the local chemical potential varies slowly in the center of the 
sample, where most atoms are residing, and the gradual character of 
the Mott transition would be less pronounced.

However, and this is the theme of the present paper, a pure Mott 
insulating state requires commensurate filling, i.~e., the number of 
particles has to match the number of wells. In a quadratic
potential 
this requirement is relaxed because excess atoms are absorbed into 
a superfluid 
region, where the filling is noninteger, at the surface of the sample. 
Using a flat-bottomed potential may put more severe constraints on 
the number of particles, which presents a problem, since in an actual 
experiment one does not have very precise control over this number. 
It is the purpose of the present study to 
explore the balance between on the one hand, the gradual character of the 
Mott transition, and on the other hand, the uncertainty in the 
particle number. 

\begin{figure}
\includegraphics[width=\columnwidth]{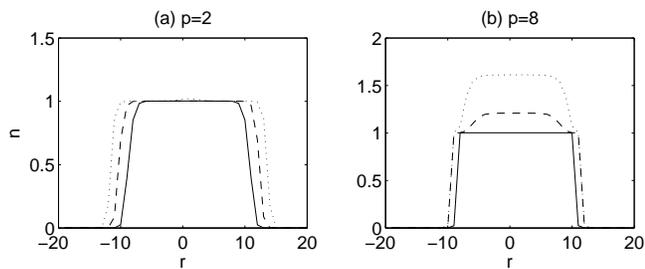}
\caption[]{Cross-section of the density profiles in a two-dimensional 
power-law potential 
with power law $p=2$ (a), and $p=8$ (b), respectively. The number of 
atoms is $N=300$ (full lines), $N=400$ (dashed), and $N=500$ (dotted).
\label{fig:densprofiles}}
\end{figure}

We consider two-dimensional systems in this study, 
since a power-law potential is more easily created in the plane than 
in three dimensions. The confinement to two dimensions can be realized 
by applying a strong optical-lattice potential in the third direction. 
The external potential is taken to be 
\beq
V(r) = (|r|/r_0)^p,
\label{potential}
\enq
where $|r|$ is the distance from the center in two dimensions 
measured in unit cells, $r_0$ is a constant length, and 
$p$ is the power law. 
In experiments with trapped atoms, quadratic potentials are prevailing 
for natural reasons (the first term in the expansion around a minimum 
of a smooth function is in general quadratic). 
A fourth-order potential has been created by superimposing a 
Gaussian optical potential on a quadratic magnetic one in order to 
observe fast-rotating 
vortex configurations \cite{bretin2004}. Other than that, realizations 
of high-order power-law potentials for atoms have been scarce. 
In two dimensions, an optical power-law potential could be realized 
using Gauss-Laguerre optical beams, Fourier optics, or masks; a 
three-dimensional trap would have to be created by combining 
several such potentials. 
Figure \ref{fig:densprofiles} shows a few examples of density profiles 
for a two-dimensional system in quadratic ($p=2$) and $p=8$ power-law 
potentials. 

The most common way of detecting the state of an atomic gas in 
an optical lattice is by time of flight, that is, to release the 
sample from the trapping potential and image it after some time 
of expansion. From the 
resulting density profile one may deduce the fraction of coherent 
atoms, i.~e., the condensate fraction \cite{stoferle2004}. 
Recently, experimental techniques have been developed for directly 
detecting Mott plateaus {\it in situ} \cite{folling2006,campbell2006}; 
nevertheless, that type of method addresses the occupation number 
and not the coherence, so the time-of-flight method remains 
the most straightforward. 
A harmonically trapped gas will in such an experiment exhibit a 
gradually increasing condensate fraction as the tunneling is increased. 

In order to observe a sharper Mott transition, it was suggested in 
Ref.\ \cite{gygi2006} that 
the potential can be made more similar to a square well. 
For a power-law potential 
of the form (\ref{potential}), the limit $p\to\infty$ corresponds to 
a square well of radius $r_0$. In a flat-bottomed potential, the 
typical situation will be that nearly all atoms will be in the same 
state (Mott insulating or superfluid), instead 
of having appreciable fractions of atoms in both states for a broad 
range of parameter values. However, now there appears a new problem: 
a flat-bottomed potential is much more sensitive to the exact number of 
particles. In a homogeneous Hubbard model, the Mott insulating state 
can only occur if the filling is commensurate, that is, if the number of 
particles is exactly an integer multiple of the number of wells: 
Adding one particle to a Mott insulator results in a superfluid. 
In a power-law potential, there is no such restriction on the number of 
particles, but the region in phase space that 
can accommodate a partially Mott insulating system will be smaller 
for a flat-bottomed potential than for a more rounded one. 
This can be seen Fig.\ \ref{fig:densprofiles}, where 
the density profiles for a one-dimensional system with different 
particle numbers are compared. In the quadratic $p=2$ trap, an 
increase of the number of particles does not have a dramatic 
effect on the profile or the number of Mott insulating atoms. 
In contrast, for a $p=8$ potential illustrated in 
Fig.\ \ref{fig:densprofiles}~(b), it is seen how an increase in 
particle number from $N=300$ to 400 takes most atoms out of the Mott phase. 

\begin{figure}
\includegraphics[width=\columnwidth]{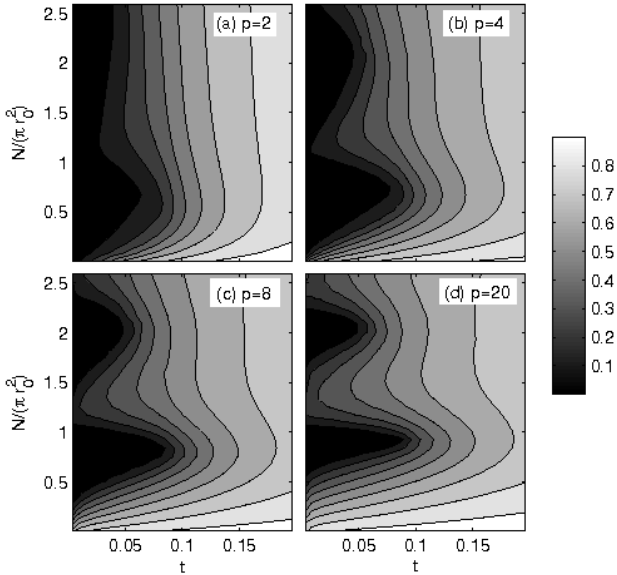}
\caption[]{Phase diagrams for a Bose-Hubbard model in power-law 
trapping potentials with power (a) $p=2$, (b) 8, (c) 20, and (d) 100. 
The phase space is spanned by the tunneling matrix element $t$ and 
the number of particles divided by the trap area, $N/\pi r_0^2$.
\label{fig:phasespace}}
\end{figure}

In order to quantify this, we 
use the Gutzwiller mean-field method with a local-density 
approximation. The fraction of Bose-Einstein 
condensed atoms is defined as 
\beq
\frac{N_C}{N} = \frac{\sum_r |\langle a_r\rangle|^2}{\sum_r \langle \ad_r a_r\rangle}.
\enq
This fraction can be inferred from experimental measurements 
\cite{stoferle2004} and is 
a direct quantitative measure of the state of the system.
Inverting the definition of a local chemical potential,
\beq
\mu(r) = \mu - V(r) = \mu - \left(\frac{|r|}{r_0}\right)^p,
\enq
one can apply a local-density approximation to transform the 
spatial sum to a sum over $\mu'\equiv\mu(r)$, which in two dimensions reads
\beq
\frac{N_C}{N} = 
\frac{\int_{\mu_0}^{\mu} d\mu' \frac{dr}{d\mu'} 2\pi r(\mu') n_C(\mu')}
{\int_{\mu_0}^{\mu} d\mu' \frac{dr}{d\mu'} 2\pi r(\mu') n(\mu')},
\enq
where $n_C(\mu)=|\langle a\rangle|^2$ and $n(\mu)=\langle \ad a\rangle$ 
are the 
condensate density and total density, respectively, for a homogeneous 
sample at chemical potential $\mu$. The lower limit $\mu_0$ is the 
chemical potential at which $N$ vanishes; it obeys $\mu_0 \leq 0$ with 
the equality holding at vanishing tunneling matrix element $t$.
These functions have been computed and tabulated, whereafter the 
integrals have been summed for a range 
of particle number $N$ and tunneling matrix element $t$.
The result 
is displayed in 
Fig.\ \ref{fig:phasespace}. 
It is seen that for more flat-bottomed potentials, the 
transition between the Mott and superfluid states, i.~e., from 
zero to non-negligible condensate fraction, is steeper. 
At the same time, the Mott state occupies a much smaller region in 
phase space for the flat-bottomed potentials. It is this tradeoff 
that has to be considered in experiment: In order to see a sharp Mott 
transition, one needs to have control over the number of particles.

\begin{figure}
\includegraphics[width=\columnwidth]{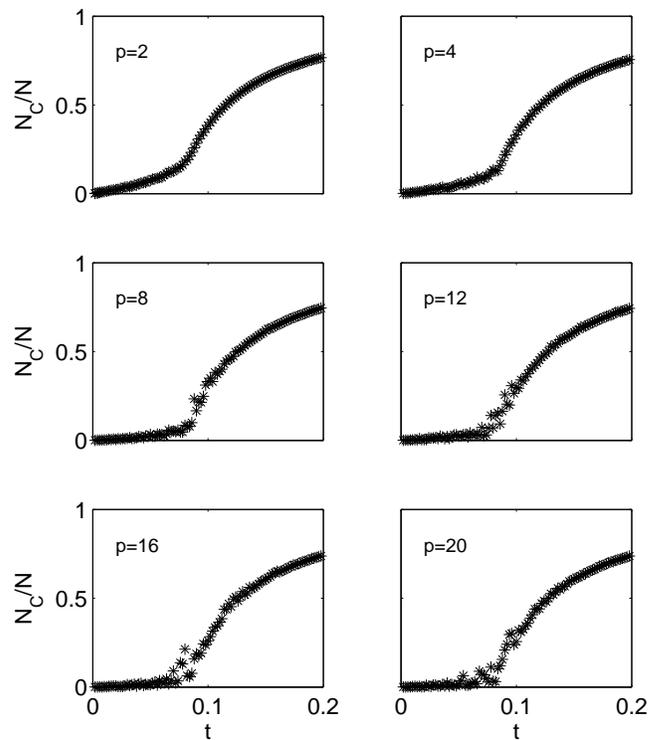}
\caption[]{Outcome of a simulated experiment measuring the fraction 
of condensed atoms $N_C/N$ in a trap, traversing the Mott transition 
as the tunneling $t$ is increased while the number of atoms is 
allowed to fluctuate within 10\% according 
to a uniform random distribution. The different curves correspond 
to different power laws $p$ of the two-dimensional trapping potential.
\label{fig:random}}
\end{figure}

Next, we simulate an experiment where the condensate fraction is 
measured for a series of increasing tunneling 
matrix element $t$ (i.~e., decreasing optical lattice irradiance), 
where $N$ has been allowed to fluctuate. 
Experimentally, controlling the number of particles is very difficult
\cite{chuu2005}, 
but the number can be measured afterwards to within certain bounds, 
so on performing a series of runs, one may apply post-selection, 
i.~e., only the data from those runs that correspond to the desired 
number is included in the analysis. 
While the relative number between different shots is quite easy to 
measure, inferring the absolute number is more problematic and may 
produce a systematic error \cite{oostenprivate}. 
In the simulation, the average particle 
number $\langle N\rangle$ is chosen to lie in the most pronounced Mott 
insulating region seen in Fig.\ \ref{fig:phasespace}. Thus, for $p=2$ 
we choose $\langle N\rangle=0.7\pi r_0^2$, and so on; for $p=20$ we 
choose $\langle N\rangle=\pi r_0^2$, corresponding to a filling of one 
particle per site. 
The fluctuations around the mean particle number are assumed to be 
uniform, as they would if the number was controlled by 
post-selection. 
Figure \ref{fig:random} shows the result for a fluctuation of 10 percent.
As expected, for the 
quadratic $p=2$ potential, the uncertainty in particle number has 
little effect, but the transition is gradual. In contrast, for the 
most flat-bottomed potential, $p=20$, the transition would have been 
much sharper, but the uncertainty in the number of particles ruins 
the observability of this sharp transition. However, 
in the intermediate range of powers, both effects may apparently be 
working in favor: the transition is 
reasonably sharp as a function of $t$, yet the Mott insulating 
state covers a broad enough range of particle number $N$ that the 
experimental uncertainty can be accommodated. Upon optical inspection, 
it seems that in this example, all power laws $p$ between 2 and 20 
yield about equally sensible results.

\begin{figure}
\includegraphics[width=\columnwidth]{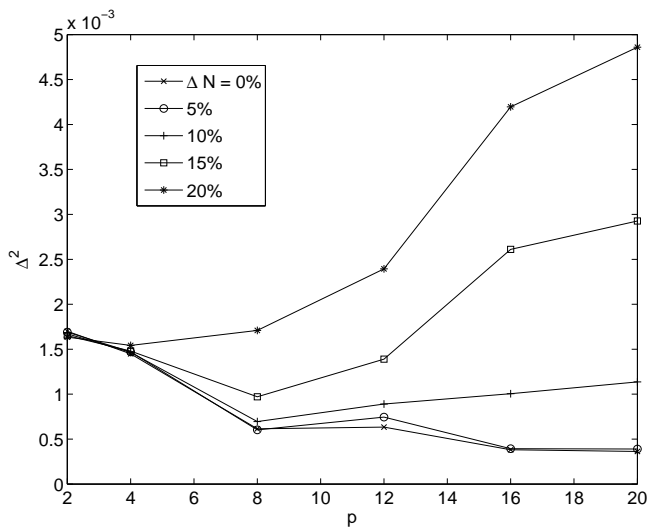}
\caption[]{Error in fitting the simulated data for the coherent 
fraction of atoms $N_C/N$, as a function of the 
power law of the trapping potential. The width $\Delta N$ of the 
simulated random distribution of particle number is indicated 
in the legend.
\label{fig:fit}}
\end{figure}

As an example calculation in order to quantify these considerations, 
we try to locate the transition by 
fitting the simulated data $N_C/N$ 
to a function of the form 
\beq
f(t) = \left\{ \begin{array}{ll}
0, & t < t_c,\\
\beta(t - t_c)^{\alpha}, & t \geq t_c,
\end{array}
\right.
\enq
where $t_c$ is a trial transition 
point and $\alpha$ and $\beta$ are fitting parameters. 
In Fig.\ \ref{fig:fit}, the squared residual 
error of the fit, 
\beq
\Delta^2 = \langle |N_C/N - f(t)|^2\rangle,
\enq
is displayed for different power laws $p$ and uncertainties in particle 
number $\Delta N$.
Clearly, the error in the fit is not dramatically different for 
different power laws, and the finite number of points included 
in the statistical sampling introduces a 
scatter in the data, but for very small uncertainty, the minimum is of 
course obtained for the most flat-bottomed trap, $p=20$. For the 
largest uncertainty of 20\%, the best fit is 
obtained for the quartic trap, $p=4$, 
where the scatter in the data is smaller. Further numerical 
experimentation, not shown here, indicates that the $p=2$ trap is 
best when the uncertainty exceeds about 30$\%$.
For the middle-ground cases, with uncertainties of 5\% and 10\%, 
it is an intermediate power law, $p=8$, that produces the best fit. 
So indeed, for a modest uncertainty in particle number there seems 
to exist a range of power law where neither of the two competing 
processes is very strong, and the clearest Mott transition occurs 
for an intermediate power-law potential.

Summing up, in power-law potentials one faces a tradeoff between two 
effects that affect the detectability of a Mott insulator. On the one 
hand, a flat-bottomed potential (i.e., a high power law) is preferred 
in order to make the Mott transition sharper. On the other hand, 
such a potential will increase the precision with which the number 
of particles has to be controlled. 
It was seen in an example simulation that 
for an uncertainty in number between 
ten and twenty percent, a power law between 4 and 8 yields the most 
accurate determination of the Mott transition; if the uncertainty can be 
made smaller, then a very flat-bottomed potential with a power law above 
10 yields the best result.

\begin{acknowledgments}
This project was financially supported by the Swedish Research Council, 
Vetenskapsr{\aa}det. 
\end{acknowledgments}



\end{document}